\documentclass[baaa]{baaa}

\usepackage[pdftex]{hyperref}
\usepackage{subfigure}
\usepackage{natbib}
\usepackage{helvet}
\usepackage[font=small]{caption}

\begin{document}


\journalvol{57}
\journalyear{2014}
\journaleditors{A.C. Rovero, C. Beaugé, L.J. Pellizza \& M. Lares}


\contriblanguage{1}


\contribtype{3}

\thematicarea{2}

\title{The Milky Way disk}


\titlerunning{The Galactic disk}


\author{Giovanni Carraro\inst{1,2}}
\authorrunning{G. Carraro}
\contact{GC: gcarraro@eso.org}

\institute{ESO, Alonso de Cordova 3107, 19001, Santiago de Chile, Chile\and
  Dipartimento di Fisica e Astronomia {\it Galileo Galilei}, Universit\'a di Padova, Vicolo Osservatorio 3,
  I-35122, Padova, Italy
}


\resumen{ 
}

\abstract{
This  review summarises the invited presentation I gave on the Milky Way disc. The idea underneath was to touch those topics that  can be considered hot nowadays in the Galactic disk 
research: the reality of the thick disk, the spiral structure of the Milky Way, and the properties of the outer Galactic disk. A lot of work has been done in recent years on these topics, but
a coherent and clear picture is still missing. Detailed studies with high quality spectroscopic  data seem to support a dual Galactic disk, with a clear separation into a thin and a thick component.  Much confusion and very discrepant ideas still exist concerning the spiral structure of the Milky Way. Our location in the disk makes it impossible to observe it, and we can only infer it. This 
process of inference is still far from being mature, and depends a lot on the selected tracers, the adopted models  and their limitations, which in many cases are neither properly accounted for, nor
pondered enough.
Finally, there are very different opinions  on the size (scale length, truncation radius)  of the Galactic disk, and on the interpretation of the observed outer disk stellar populations in terms  either 
of external entities (Monoceros, Triangulus-Andromeda, Canis Major), or as manifestations of genuine disk properties (e.g., warp and flare).} 




\maketitle

\section{Introducción}
\label{S_intro}

Being the galaxy we live in, the Milky Way has always attracted a lot of attention. 
The last couple of decades have been dominated by
the concept of astronomical survey, and well-known examples are 2MASS, WISE, SDSS/SEGUE, APOGEE, and many others. These surveys, and the future ones, 
are generating  enormous amount of data, and are
changing the way data are stored, reduced, and analysed. 
Surely, this wealth of data will contribute significantly to our knowledge of the Milky Way, its structure, chemical and dynamical evolution, being the ultimate
goal to understand how our galaxy formed and how it was assembled.\\

\noindent
From our location, at about
8.5 kpc from the Galactic Center, we can obtain precise information -both from ground and with dedicated satellites- for many individual stars, star clusters, and molecular clouds.
However, this does not represent always a real advantage when we try do derive general properties of galactic components, e.g. the bulge, the halo
and the disk, with their various substructures. 
We do not have the same clean global view of the Milky Way that we have for external face-on galaxies.
The Sun position, inside  the Orion spiral arm (also called Local arm or Orion spur), makes it challenging to disentangle structures that
accumulate along any line of sight and to position them precisely enough.\\

\noindent
In this review, I will try to summarise in a critical manner the present day understanding of three particularly hot topics in modern Galactic structure.\\

\noindent
First of all, I will address the issue of the reality of the Galactic thick disk as a separate entity, by comparing the outcomes of large surveys with the results of smaller-size, dedicated projects.
In several cases, small scale projects, but dedicated, provide cleaner results than large-scale, limited precision, surveys.  Cheer numbers do not compensate for poor data.\\

\noindent
Second, I will touch the classical topic of the spiral structure of the Milky Way. We do not observe the spiral structure in the Galactic disk, but we need to infer it.
Different tracers provide different pictures of the Milky Way spiral structure.
Charting the Milky Way still requires heavy model assumptions, to derive distances from kinematics  or from magnitudes, to account for extinction, sample contamination, 
and so forth.\\

\noindent
Third, I will discuss the structure of the outer disk, the anti-center. Is our disk truncated? Is there an edge, or a cut-off in light or mass? Is the anti-center 
one of the main arenas where the accretion of external material in the form of dwarf galaxies and subsequent formation of streams is taking place? Or, more conservatively,
is what we observe the manifestation of the intrinsic structure of the disk? Structures like the Galactic warp and flare have not always been correctly accounted for in the interpretation of the stellar population in the outer Galactic disk.

\section{Question \# 1:  {The Galactic thick disk}: it is more effective to have a large, low quality, or a small, high quality, sample?}
The paradigm that the Milky Way disk is dual, originally proposed by \citep{GilRei83} has become recently matter of intense discussion. The separation into thin and thick disk
was found by counting stars toward the south Galactic pole. The density law, derived from the luminosity function of stars fainter than $M_V \sim$ +4 was found to follow a single exponential
with scale height $\sim$ 300 pc below 1000 pc from the disk plane, and a second exponential with scale heigh $\sim$ 1450 pc above 1000 pc (see Fig.~1). \cite{GilRei83} called the first structure
the thin disk, and the second the thick disk. \\

\begin{figure}[!ht]
  \centering
  \includegraphics[width=0.45\textwidth]{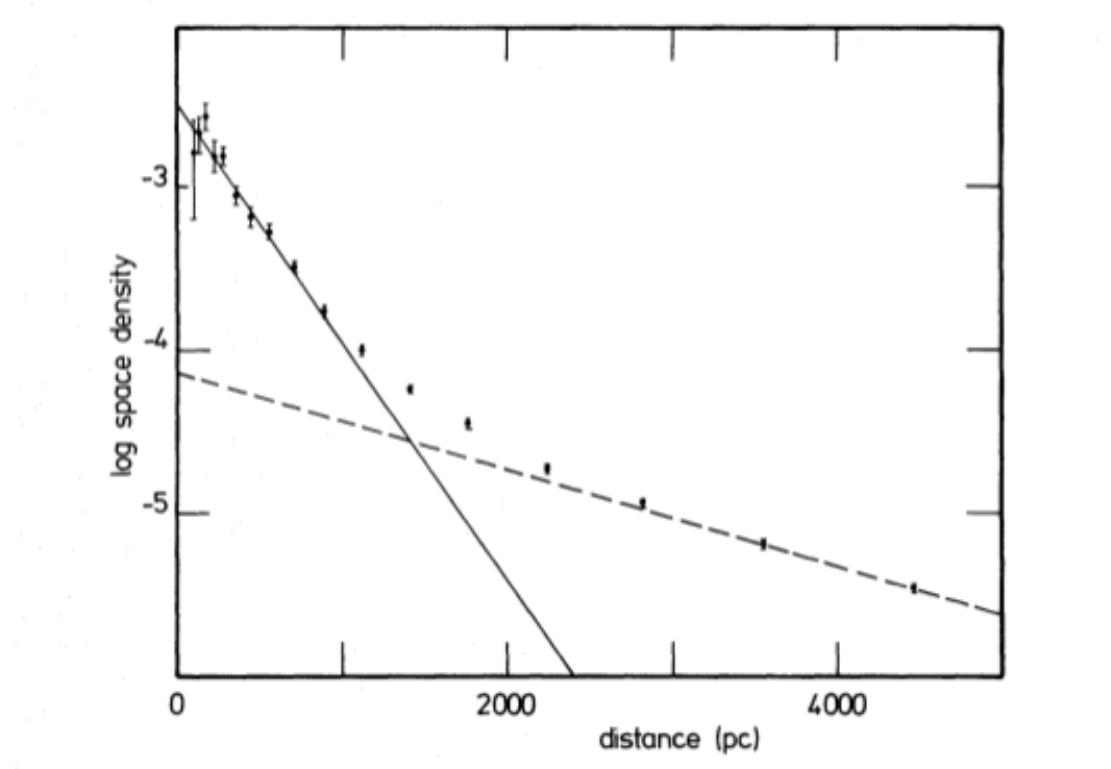}
  \caption{Star counts toward the South Galactic pole as a function of the distance from the Galactic plane, from \citep{GilRei83}}
  \label{F_toobig}
\end{figure}

\begin{figure}[!ht]
  \centering
  \includegraphics[width=0.45\textwidth]{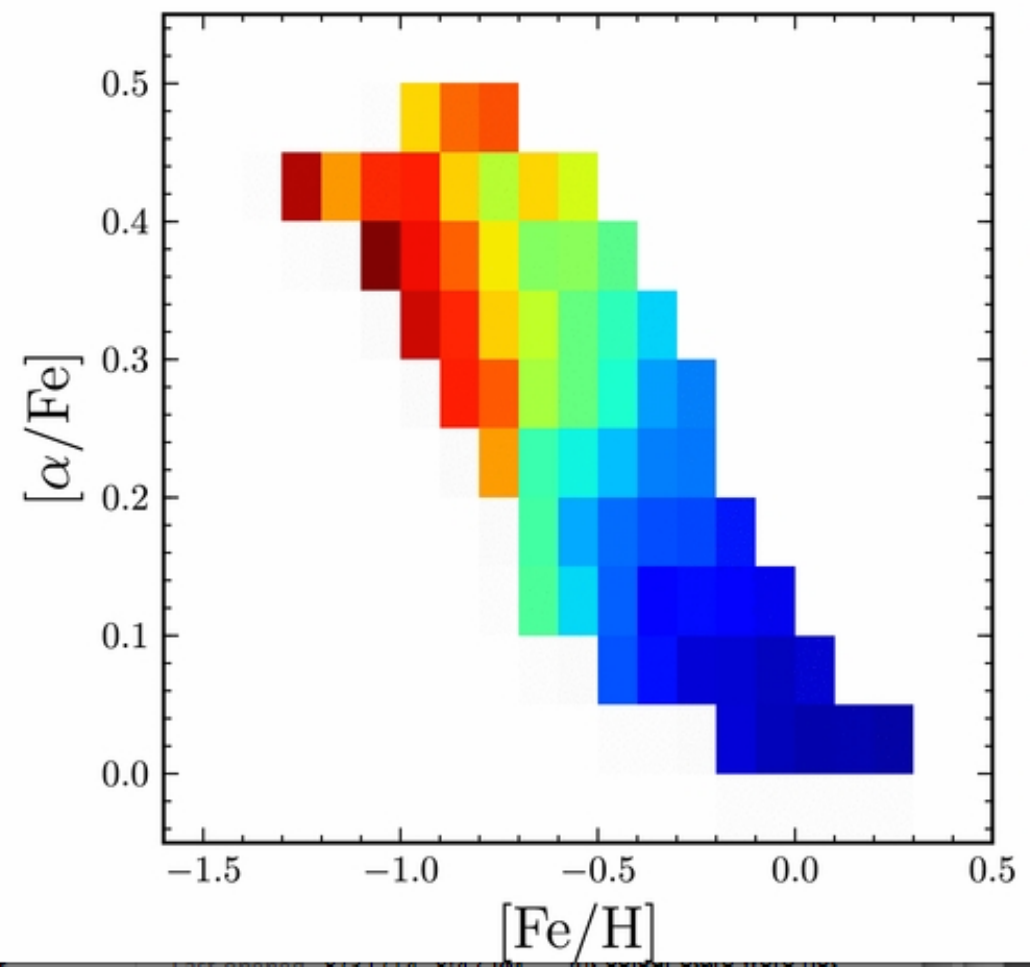}
  \caption{The concept of mono population box. Each pixel represents the average chemical properties of at least 100 stars from SDSS/SEGUE.}
  \label{F_toobig}
\end{figure}

\noindent
This paradigm resisted for about 30 years, with variations only in the values of the two scale heights.  However, recently, \citep{Bovy12a} and \citep{Bovy12b} questioned this paradigm and proposed
that the Galactic disk has an unique vertical  scale length. Whether the disk possesses one or two scale lengths has crucial implication for our understanding on how the disk built up, and therefore
\citep{Bovy12b} claim had quite an impact and generated much discussion.\\
 
\noindent
\cite{Bovy12a} based their claim on the analysis  of $\sim 24,000$ stars from the SDSS/SEGUE survey, for which radial velocity, distance, and basic abundance analysis was available on an individual star basis. The typical uncertainty in metallicty [Fe/H] and [$\alpha$/Fe] ratio that one can obtain with these low resolution ($R\sim2,000$)
spectra is 0.2 and 0.1 dex, respectively. \citep{Bovy12a} introduced the concept of mono abundance stellar populations, which uses the [$\alpha$/Fe] ratio as a proxy for the age of a stellar population. Ages for individual stars are in fact difficult to estimate using such low resolution spectra. By separating the whole sample into statistically significant  mono populations boxes, as illustrated in Fig.~2, \citep{Bovy12b} found a continuum in chemical properties as a function of vertical distance from the plane, concluding that the disk exhibits a smooth variation in its population properties moving away from the formal Galactic plane. Notice that the size of a pixel is 0.05 dex in  [$\alpha$/Fe], and 0.1 in [Fe/H].

This approach is obviously subject to criticisms, because  population properties cannot be efficiently pinned down using such low resolution data, and this limitations cannot be  any means can be compensated or alleviated by the large number of stars used. 
The large typical uncertainties in chemical properties clearly have the effect of smearing stars across boxes in a unpredictable way.\\

\noindent
A different approach has been adopted by \citep{Bensby14}.  They derived full abundance analysis for a sample of {\it only} 740 F and G dwarfs out of spectra with resolution in the range R=40,000-110,000 and high signal-to-noise ratio (150-300).  The sample was carefully built up to trace  stellar populations along a wide range in metallicity, which presumably cover the whole
disk stellar populations. The strength of this approach if that the uncertainties in  the properties of individual stars are virtually negligible,  it is possible to derive solid estimates
of star's ages, and it becomes easy to separate efficiently stellar populations, when they actually differ.\\

\begin{figure}[!ht]
  \centering
  
  \includegraphics[width=0.45\textwidth]{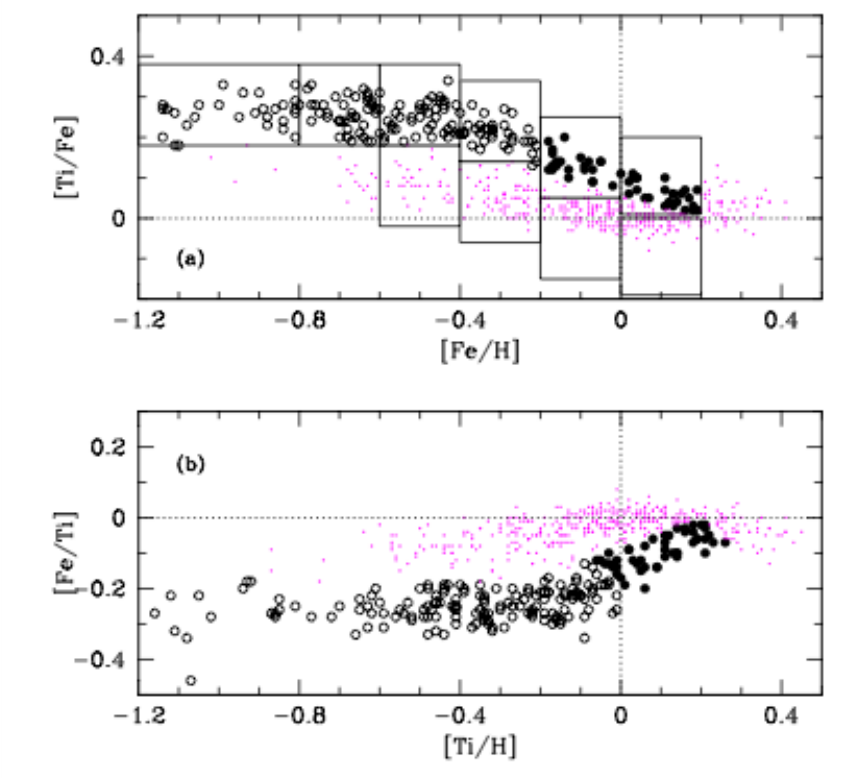}
  \caption{Individual thin (purple symbols) and thick (open circles) disk star chemical abundances from \citep{Bensby14}.}
  \label{F_toobig}
\end{figure}

\noindent
As an illustration, Fig.~3 shows the stars distribution in the abundance plane, similar to Fig.~1. The only difference is that in this cases individual stars with negligible uncertainties are plotted instead of mono-population boxes. 
$\alpha$-enhanced, metal poor stars  (open circles) neatly separates from moderately/solar $\alpha$-enhanced metal rich stars (purple symbols). 
These latter stars are considered to trace the Galactic thin disk, while the formers trace the Galactic thick disk.
The separation occurs at [Fe/H] $\sim$ -0.7, which is therefore the lowest thin disk metallicity. There is also a group of intermediate properties stars, with high metallicity and 
moderated enhancement in 
$\alpha$ elements (filled circles).  These stars would represent the metal rich tail of the thick disk. This scenario is confirmed by the early results of the Gaia-ESO survey (\citep{Miko14}).\\

\noindent
In this scenario, which data seem to support, thin and thick disk are therefore two separate entities.  This rules out an internal  formation scenario, in which the disk thickens because of 
dynamical heating,
and old, metal poor, $\alpha$-enhanced stars occupy higher and higher regions of the Galactic disk. Instead, an accretion scenario seems to be more plausible for the origin of the thick disk 
\citep{Bensby14}.

\section{Question \# 2:  The spiral structure of the Milky Way : is there an ideal spiral structure tracer?}
The early history of the spiral structure of the Milky Way is a fascinating one. \citep{Alex852} was the first to mention that the Milky Way is probably a spiral galaxy. 
Other authors
made this suggestion afterwards, based on the observation of external spirals, and the existence in the Milky Way of the same components (gas, dust, young stars...) that are
present in spiral galaxies.
The real quest for the spiral structure of the Milky Way  started after the first world war only. Kapteyn, at Leiden laboratories, attempted at discovering the spiral structure of the Milky Way by counting stars on the ground that spiral arms are over-densities of stars. 
This attempt failed for a variety of reasons: photographic plates were not particularly efficient, the absolute magnitude of different-colour (i.e. spectral type) stars were not known, and it was not possible to invert the equations for star counts to derive distances.\\

\noindent
It was during the second world war that a huge step-ahead was done by W. Baade. He made use of the Mt. Wilson telescope during the Los Angeles black out to map the disk of M31 in search of  HII regions and  discovered that HII regions are not randomly distributed across M31 disk, but they form a large scale  structure which resembles a spiral 
galaxy, with rings, bridges, bifurcations, and so forth.
Whether these structures can be identified as  logarithmic spirals or no, it is very hard to say, as illustrated in Fig.~4, taken from \citep{Liszt85}.

\begin{figure}[!ht]
  \centering
  
  \includegraphics[width=0.45\textwidth]{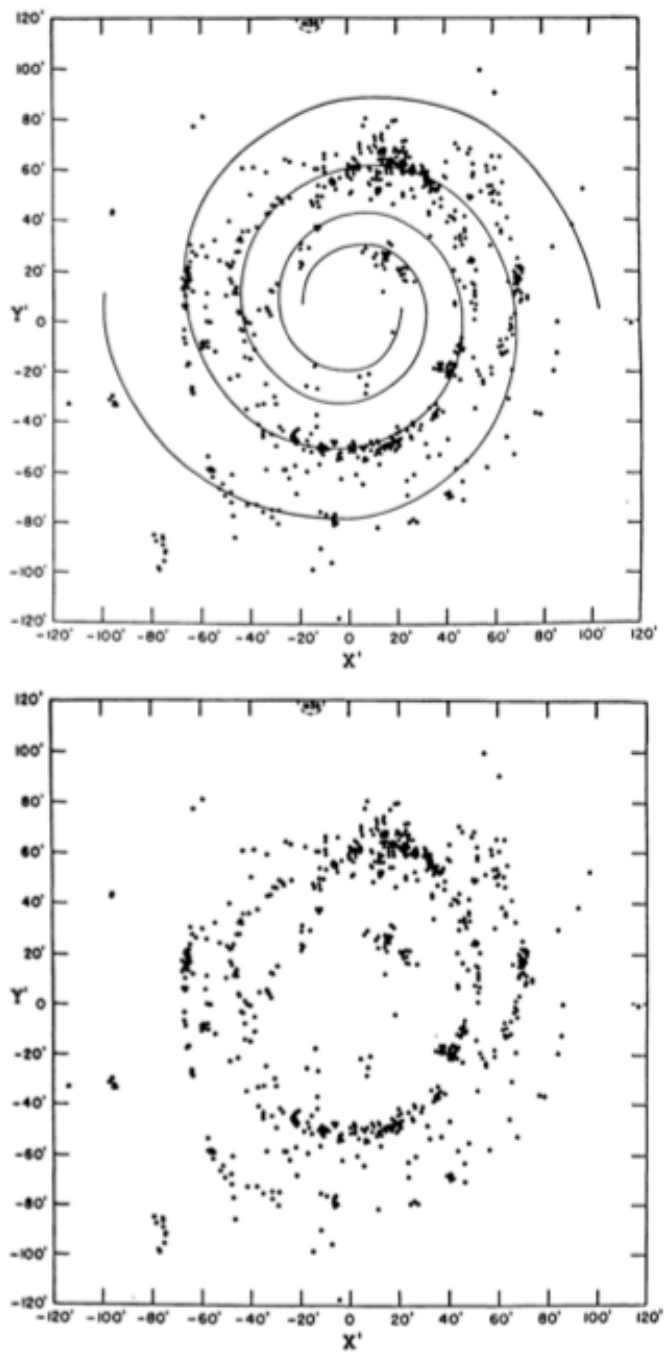}
  \caption{HII region distribution in M~31 from Baade (lower panel), fitted with logarithmic spiral arms (top panel)}
  \label{F_toobig}
\end{figure}

\noindent
M~31 is almost face-on and this would make it easy to distinguish spiral features.  Our incapability to describe clearly  in whatever mathematical model M~31 spiral structure indicates
how difficult it can turn when we look at our own disk.\\

\noindent
After Baade work on M~31, two parallel lines of investigations opened. \\

\noindent
On one side, the recent discovery of the 21cm line boosted the search for HI emission/absorption in the Galaxy and  the 
compilation of the first HI map of the inner Milky Way disk (the
Sydney-Leiden map, see Fig.~5, \citep{Kerr62}).  This map was laid down using the Oort results for the Galactic disk kinematics in the inner Galaxy, where objects are assumed to move in circular orbits. In this case,
a distance can be derived from radial velocity, assuming that this radial velocity is  caused by a real cloud (either HI or CO) moving inside an arm, and not by whatever arm internal dynamical mechanism, like for instance streaming motions. The same Oort formulae cannot be used outside the solar ring, and other independent distance indicators have to be adopted.
Besides this, other  limitations are that our galaxy is filled with HI, and apparently inter-arm regions have the same physical size as spiral arms. It is therefore  extremely difficult to disentangle features when they overlap in velocity along the line of sight.
In spite of all these limitations, HI and CO surveys continued to produce data in an industrial manner until now.\\

\noindent
On the other side, following up Baade results, Walter Morgan searched for blue, bright stars in the Sun vicinity, and made use of the spectral classification method 
he developed  to derive distances for stars of O and B type. The first Milky Way spiral structure sketch from this handcrafted project was published in \citep{Morgan52} for about 30 stars (see Fig.~6).\\

\begin{figure}[!ht]
  \centering
  
  \includegraphics[width=0.45\textwidth]{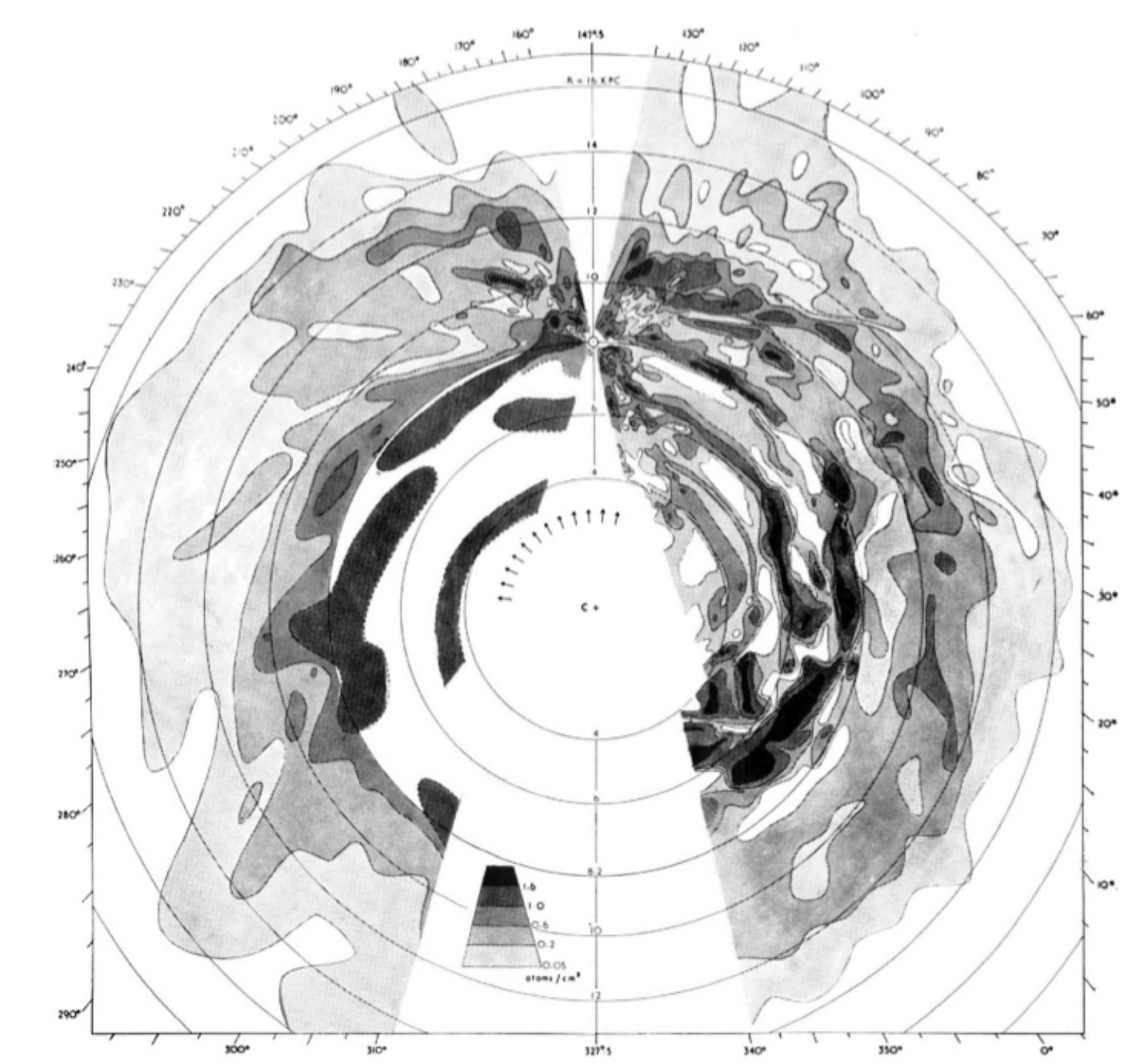}
  \caption{The Leiden-Sydney HI map of the Milky Way from \citep{Kerr62}.}
  \label{F_toobig}
\end{figure}

\begin{figure}[!ht]
  \centering
  
  \includegraphics[width=0.45\textwidth]{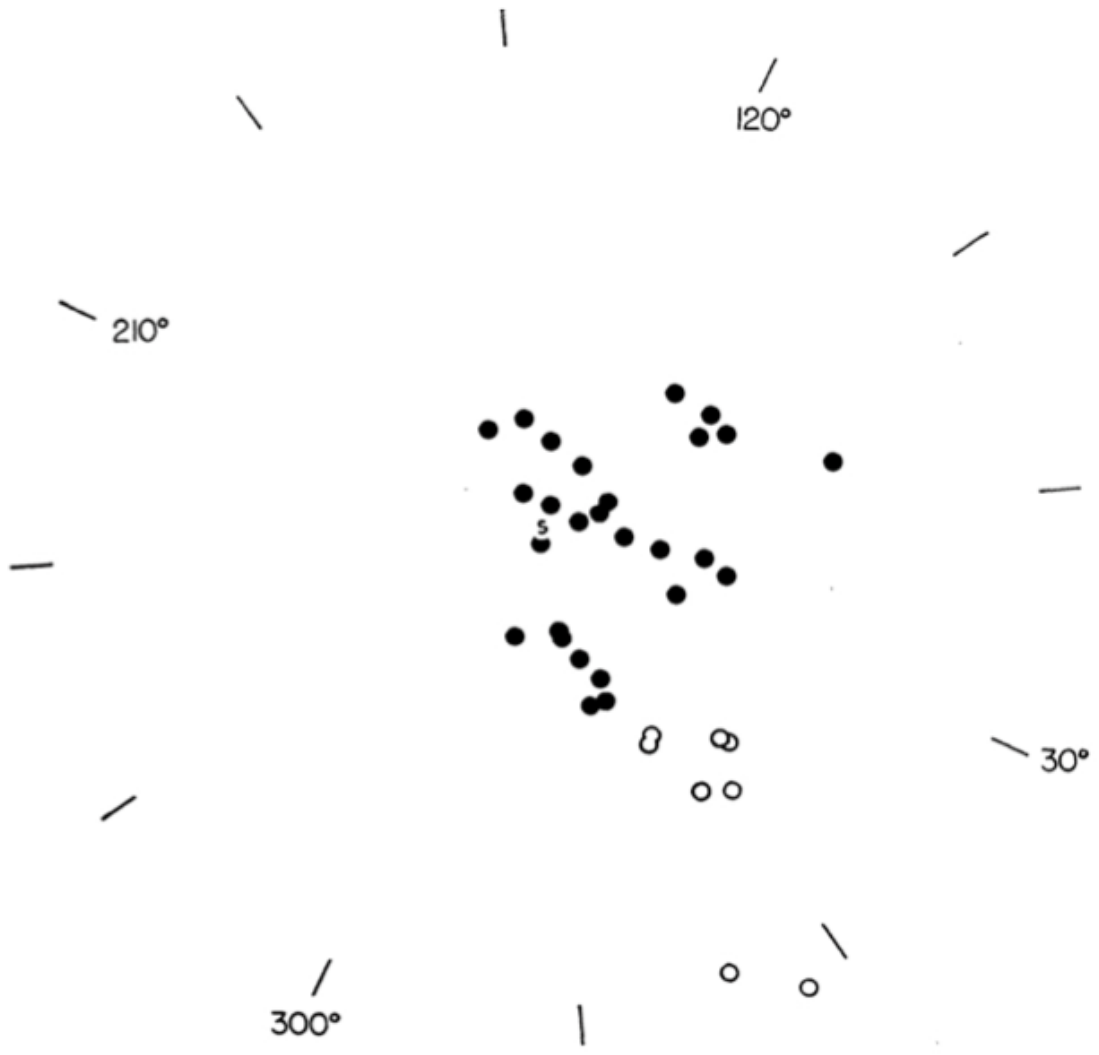}
  \caption{The spiral structure of the Milky Way from \citep{Morgan52}.}
  \label{F_toobig}
\end{figure}

\noindent
At that time it was extremely difficult to converge on a clear description of the Milky Way spiral structure.\\

\noindent
The accumulation of data in the last 60 years or so did not improve much this picture.  There is a general consensus that the Milky Way spiral structure is pretty much as
depicted in the artistic rendered realisation from \citep{Wise09} (see Fig.~7). This has been drawn counting red clump stars from the GLIMPSE survey, and filtering star counts
with HI and CO data. Despite it was intended to be a realisation for the general public, this has become a sort of widely accepted modern view of the Milky Way.
According to this picture, the Milky Way is a two-arm spiral. The two major arms are the Perseus arm  and the Scutum-Crux-Centaurs arm. There are also two minor arms, the Carina-Sagittarius and the Norma-Cygnus (or outer) arm. This picture of the Milky Way as a two-arm spiral is the classical picture proposed since the earliest times by the radio-astronomy community,
but disagrees significantly with the picture proposed by the optical community. OB stars, HII regions and young star clusters indicated that Carina-Sagittarius is one of the strongest spiral
arm in the Milky Way. The same seems to be true for the outer arm. In this view, the Milky Way would be more probably a four-arm spiral \citep{Russ03}.\\

\begin{figure}[!ht]
  \centering
  
  \includegraphics[width=0.45\textwidth]{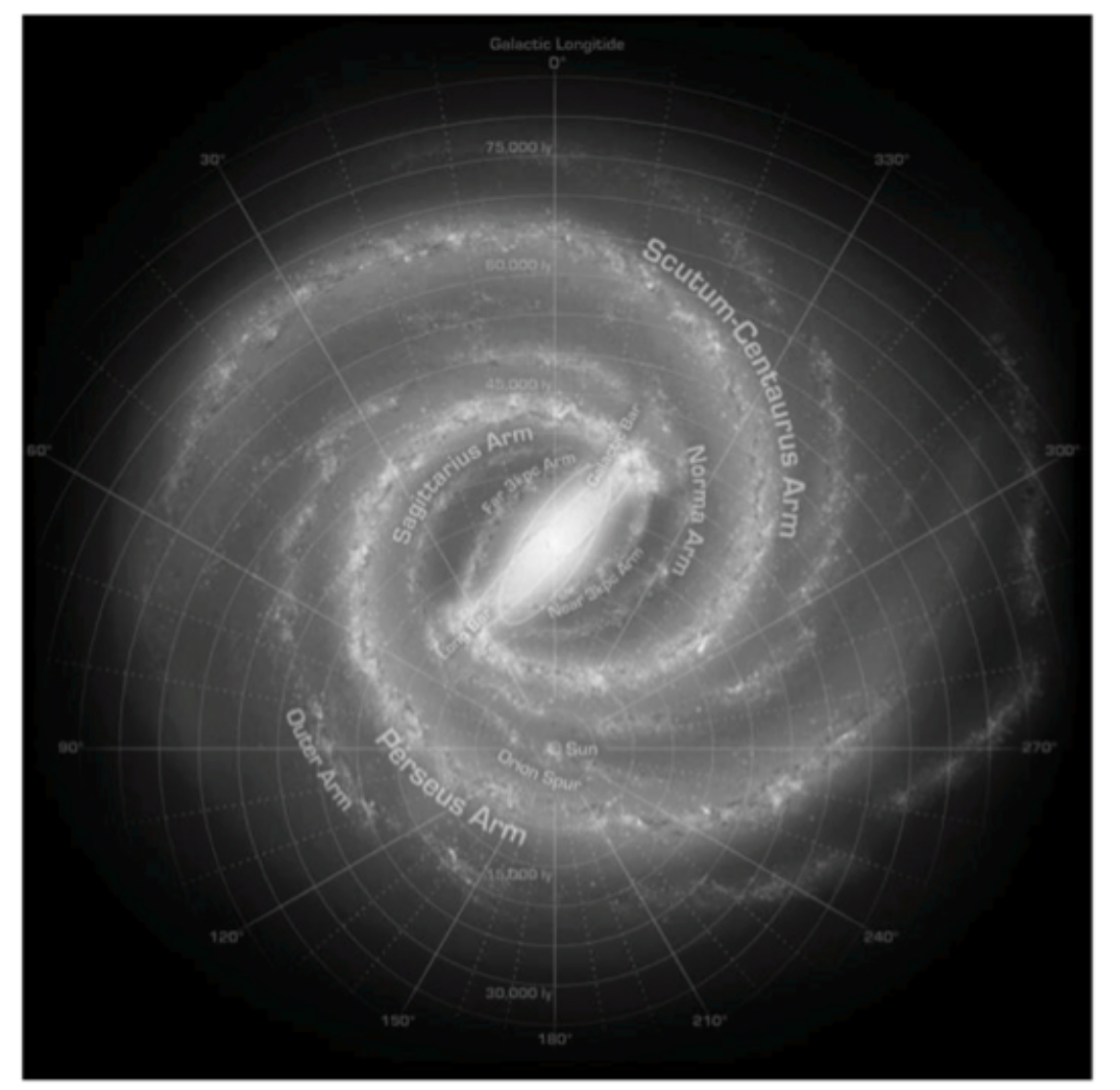}
  \caption{Artistic rendered picture of the Milky Way spiral structure from \citep{Wise09}.}
  \label{F_toobig}
\end{figure}

\noindent
As I will stress it again later on, the use of red clump stars, say stars that are burning He in their core, is tricky. Surely these are well-known distance indicators, 
because their absolute magnitude depends very mildly on age and metallicity. However, clump stars span a wide range in age (from the Hyades to 47 Tuc) and therefore
trace a variety of different stellar populations, not necessarily associated with the young, gaseous, dusty spiral arms. Besides, selecting clump stars in real life might be quite a cumbersome
task, because of the huge and unpredictable contamination from red giant branch stars, sub-giant branch stars, and reddened nearby dwarfs along any line of sight.\\
The WISE view of the Milky Way spiral structure contradicts recent findings regarding  the structure of the Local  (Orion) Arm and spiral structure in the third quadrant 
(\citep{Ruben08}), and the detailed distribution of HII regions in the first Galactic quadrant \citep{And12}.\\

\noindent
To illustrate how embarrassing the actual situation is, I will focus only on the Local arm, the closest to us, the one  which the Sun is located in. \citep{Morgan52} originally referred
to this spiral feature as a spur, a bifurcation departing from the Sagittarius arm in the first quadrant, close to the huge HII region W51 
\citep{W51}, and displacing toward Perseus in the third quadrant.
On the opposite, in \citep{Wise09} the Local arm appears as a truncated normal arm, floating in between and almost parallel to the Carina-Sagittarius and Perseus arms. Such view
is supported by the  recent maser study by \citep{Xu13}. These authors used maser data a couple of kpc around the Sun to conclude that the Local arm resembles a gran design spiral
arm both spatially and kinematically.
Strong support to the idea that the Local arm might be a spur, or a bridge is given in \citep{Ruben08}. Using a large sample of young open clusters, and young stellar population in the background, \citep{Ruben08} found that the Local arm extends into the third Galactic quadrant all the way to the Outer Norma Cygnus arm, breaking the Perseus arm, in full similarity
with what one can see in the giant spiral galaxy M~74.\\

\noindent
This plethora of different interpretations for just the closest-to-us spiral arm witnesses how much work still needs to be done, and how communication and 
critical, open-minded, sharing of results among different studies is needed. The expectations from the Gaia satellite are enormous with respect to the structure of the Milky Way in the solar vicinity.

\section{Question \# 3: The outer Galactic disk: do we have a realistic model of the Milky Way? }
The outer disk of the Milky Way has been undergoing a renaissance of interest since about 10 years.  There are essentially two reasons for that.
First, in the anti-center the stellar density is low, and the outer disk merges with the halo. This makes the disk outskirts the ideal location
where to search for over-densities which can indicate past or ongoing accretion events.
Second, the reconstruction of the radial light/mass density profile along the disk allows us to determine whether or not the Galactic disk has a density break in the anti-center, and where this break is placed. This would help us to understand which class of spiral galaxy the Milky Way belongs to (type I, II, or III, \citep{Lai14}). This is inferred by deriving the ratio of the break position to  the bar scale length.\\
Both topics have in the past made extensive use of Galactic models, like the Besan\c{c}on \citep{Robin89} or Trilegal \citep{Girardi05}, 
to predict star counts in a given Galactic direction, and to compare them with actual observations.

\subsection{The Galactic warp and flare}
Two important features that have not always been properly taken into account when studying the outer disk are the warp and the flare.
The warp of the disk was first detected in HI (e.g. \citep{May93}), and then also found in the stellar component \citep{CarSei93}.  A comprehensive summary of the warp properties is provided in \citep{Yaz06}. The maximum of the warp in the out disk occurs at longitude $\sim$ 245 deg in the third Galactic quadrant. The effect of the warp is to bend the disk down from its formal plane (at $l=0^{o}$) in the third Galactic quadrant and bend it up in the second quadrant. The line of nodes is at l $\approx$ 160, not precisely toward the anti-center direction. All traces employed so far to probe the outer disk consistently reproduce
these properties.
\cite{Yaz06} also mention the flaring of the outer disk, which is mostly seen in intermediate age or old stellar populations (pulsars, red clump and red giant branch stars; see also \citep{Corre14} and \citep{Kal14}). 
The flaring consists of an increase of the vertical scale length at increasing distance from the Galactic center. 
Recently, the flaring has been detected also in young stellar population, like HII regions \citep{And14}, cepheids \citep{Feast14}, \citep{Chakra15}, and young stellar clusters \citep{Car15}.

\begin{figure}[!ht]
  \centering
  
  \includegraphics[width=0.45\textwidth]{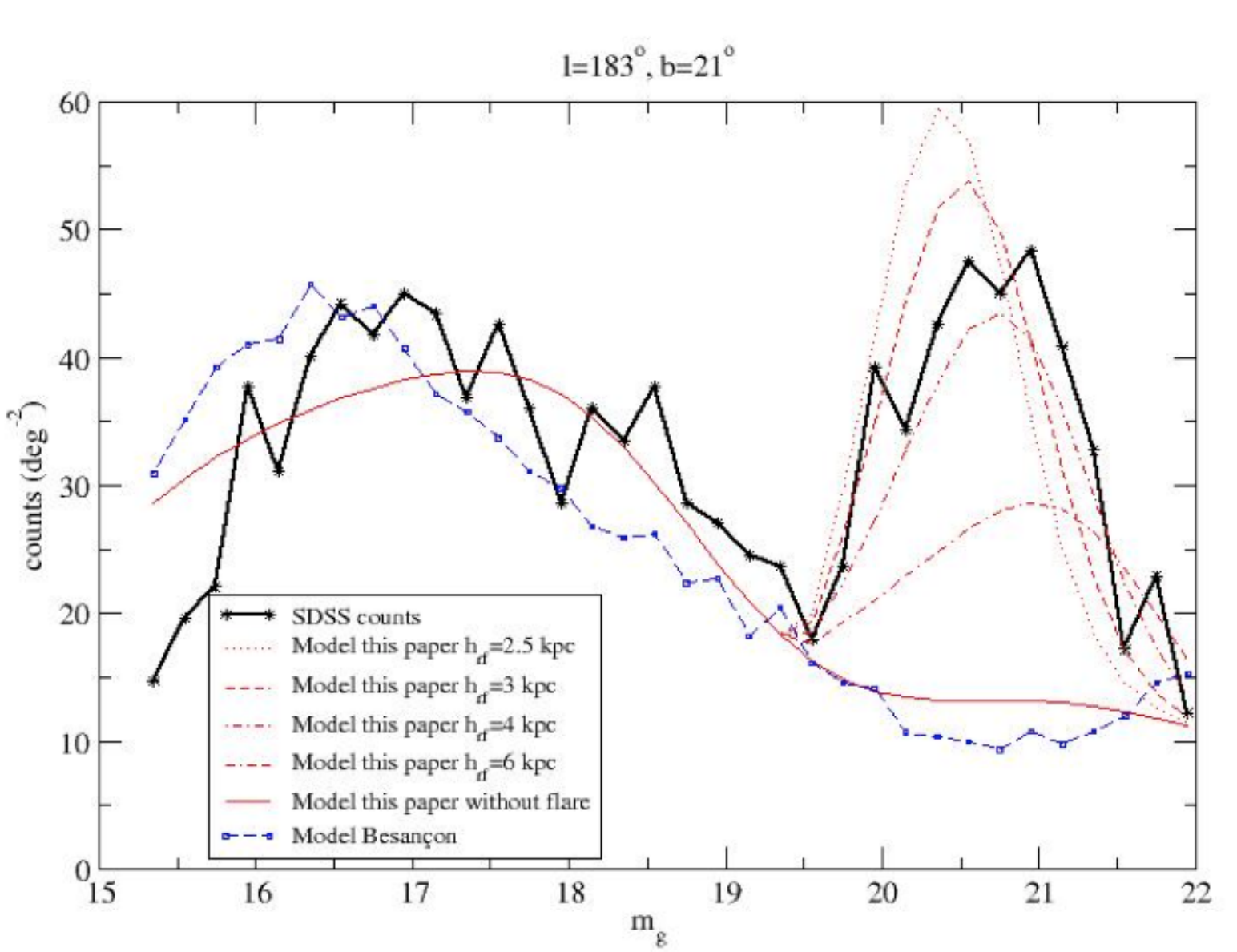}
  \caption{SDSS star counts in the anti-center, with super-imposed model accounting for the stellar flare.}
  \label{F_toobig}
\end{figure}

\subsection{Satellites and Streams in the outer disk}
Over the years, several over-densities detected in the outer disk  have been claimed to be satellites or streams in the Milky Way.\\

\noindent
The most notorious satellite was identified as an over density of red clump stars toward the Canis Major (CMa) constellation, and named the CMa dwarf galaxy \citep{Mar04}, at 
$l=244^{o}$, $b=-8^{o}$ (nowadays is best referred to as the CMa over density) .  It was identified as a probable satellite because of the presence of both an intermediate-age and a young stellar population (this latter called the blue plume). Since the Besan\c{c}on model could not reproduce this feature in the Color Magnitude Diagram of the expected Galactic population along this line of sight, it was concluded that Canis Major would be the closest dwarf spheroidal of the Milky Way, and that the blue plume would represent its last star formation episode.\\
However, the Besan\c{c}on model at that time had hardcoded an artificial cut-off of the disk at 12 kpc from the Galactic center, and did not include
a prescription for the Galactic warp and flare. 
Follow up studies on the putative young stellar population in CMa demonstrated that its more plausibly related to the spiral structure beyond the solar ring \citep{Moi06}. The chemical and age properties of the intermediate age population are within the expected ones for the Galactic thick disk, and no old stellar
populations has been detected \citep{Yaz06}, which is instead  ubiquitous among dwarf galaxies in the Local Group.\\

\noindent
CMa was believed to be for a while the core of a dwarf galaxy engulfing the Milky Way along an in-plane orbit, and its tidal stream was identified with the so-called Monoceros Ring 
\citep{New02}. This is an over density, visible both in the southern and northern Milky Way disk, virtually encompassing the whole disk.
When the association of Monoceros with CMa was proved wrong, the scenario for Monoceros moved to the idea that is would be the left over of a tidally disrupted galaxy. As in the case of CMa, neither old stellar populations has been found in Monoceros, nor young stellar population.
The age and metallicity distribution among Monoceros stars resembles very closely the intermediate-age Milky Way thick disk.
In fact, an alternative explanation has been proposed for the Monoceros ring, which would simply be produced by the flare in the outer thick disk (see Fig.~8).
Again, the Besan\c{c}on model does not include any realistic modelling of the flare, and therefore does not predict the Monoceros Ring. Once a reasonable
model of the flare is introduced, star counts in the anti-center are well reproduced \citep{Corre14}.\\

\noindent
Finally, I would like to mention another outer disk over-density, the Triangulus Andromeda (TriAnd) feature, which is seen in the second Galactic quadrant, at b $\sim$ +30 degr, right behind the Monoceros Ring \citep{She14}. TriAnd is a loose, extended structure, visible with M giants, whose metallicity [Fe/H] is around -1.2 .
Since there is no obvious center, this over density is considered as a disrupted dwarf, as in the case of Monoceros.
This interpretation has never been challenged so far. However, in a scenario where the  disk is flared, TriAnd, located  in the background of Monoceros, could as well be  Galactic 
thick disk stellar population.

\begin{figure}[!ht]
  \centering
  
  \includegraphics[width=0.45\textwidth]{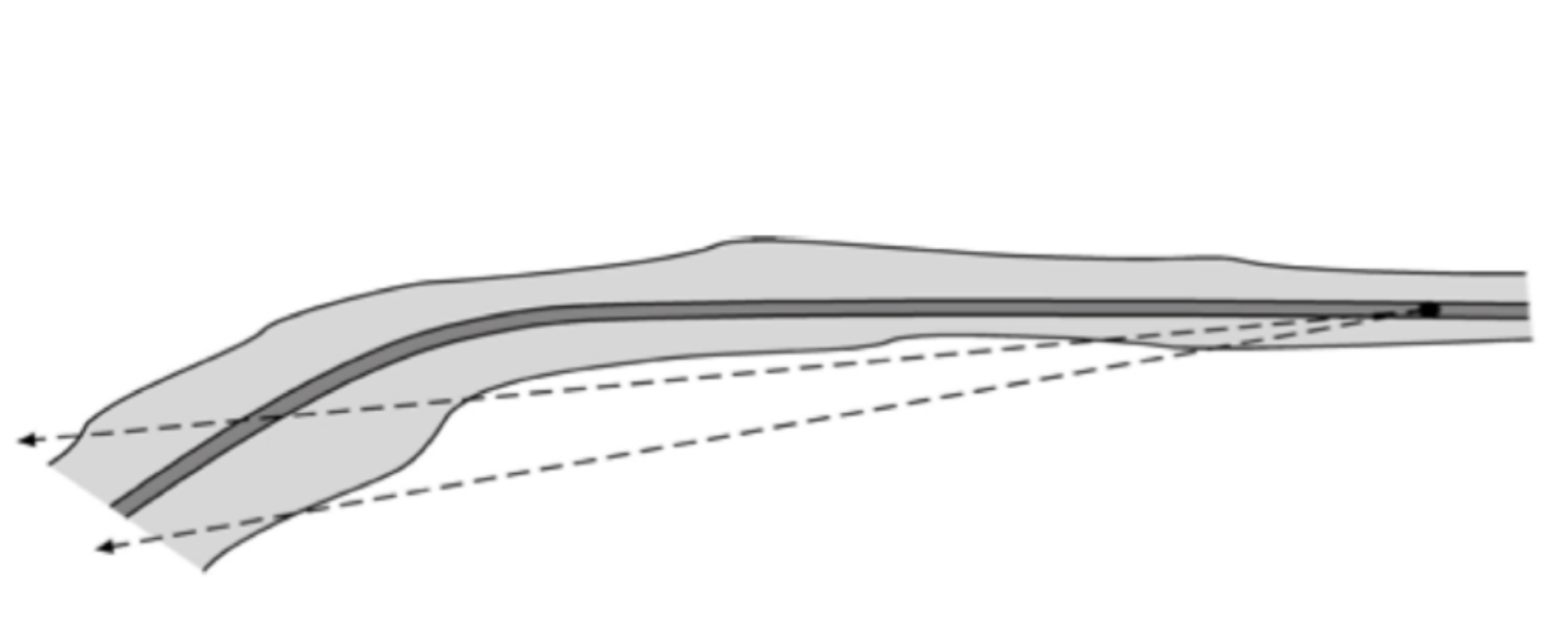}
  \caption{Visualization of the warp structure toward the outer disk.}
  \label{F_toobig}
\end{figure}

\subsection{Does the Galactic disk have a break?}
Breaks in stellar density have been found in many spiral galaxies \citep{Lai14}. When a break is found, the spiral is defined as a type II. Type II spirals are
the most common. Detecting a break in the Milky Way would help us to understand which kind of spiral galaxy we live in.\\

\noindent
\cite{Robin92} first observed A type stars toward the anti-center, and found a sudden density drop at 12 kpc from the Galactic center, which they called edge of the disk. Some caution has to be used here. Terms like edge, truncation, or cut-off refer indeed to sharp density drop, and are not the right 
description of what we mean with break. A break is a change of slope in the density gradient of the disk. Breaks can be down-bending if they correspond to a density decrease, or up-bending if they correspond to a density increase.
What \citep{Robin92} found was an artificial cut-off  (see Fig.~9 for an illustration of the effect) caused by the down-bending of the disk because of the warp and the decrease of the star density along the formal plane caused by the flare. Since their Galactic model predicted more stars than observed, a cut-off was included. More recent studies (\cite{Car10}) show that in fact the disk extends much further.

\cite{Sale09} used a different strategy. They counted A type stars from IPHAS in the sector $160\leq l \leq200$ and $-1\leq b \leq+1$ . This selection in galactic latitude
is unfortunate because the effects of both the flare and the disk are lost.Then they used the Besan\c{c}on model imposing that star counts had to be reproduced by a double exponential function, and found the best fit double exponential for a break at R =13$\pm$1.1.  

\cite{Minni11} used clump stars from the VVV survey and found an edge (not a break !) at about 12-14 kpc, depending on the direction.
They selected stars in the b range -2,+2, again an unfortunate selection if one wants to take warp and flare into account. Clump stars were selected because they are good distance candles. However, selection effects and contamination can be a killing factor. In fact, clump stars in typical 
Galactic disk color magnitude diagram are blurred by reddening, errors, and age/metallicity effects, which make them to be easily confused with red giant branch, sub-giant and nearby dwarf stars. Besides, clump stars have ages from about half a Gyr to 12 Gyrs, and therefore they may represent quite
different stellar populations (thin or thick disk, halo, or even the bulge/bar). The handling of these effects is in general very poor.\\
The same type of  stars are being used by Robert Benjamin, although in a different wavelength  regime (the mid infrared) from GLIMPE/WISE. The results have never been published so far, but it seems that a break would be present in the Milky Way at about 13.5 kpc from the center.

\section{Conclusions}
Deciphering the structure and understanding the chemical and dynamical evolution of the Milky Way disk are among  the hottest topics in modern astrophysics. 
In this review, I discussed the status of present day knowledge of three topics: the thick disk, the spiral structure of the Milky Way, and the properties of the outer disk.
Clearly there are no general consensus, and much work still needs to be done.  This is particularly relevant for the spiral structure of the Milky Way, which is often deemed
to be understood. However, even the Orion arm, where the Sun is located in, nature is controversial. The stellar and gaseous disks exhibit a break outside coronation, although
the  exact location is still disputed. The disk has a break, but not an edge, nor a sharp cut-off.  Finally, recent high quality spectroscopic data lend support to a dual disk, with a thick disk
well separated from the thin disk, both in chemical and spatial properties.\\

\noindent
If GAIA will fulfil its promises, in a time scale of half a decade or so me might get answers for several of the still open questions.

\begin{acknowledgement}
I wish to thank deeply the organisers for inviting me to this great conference. I am much indebted to T. Bensby, D. Gadotti,  J.-C Mu\~{n}oz-Mateos,  B. Burton, and Y. Momany for
many useful and inspiring  discussions. I am also grateful to the colleagues who gave me the permission to use in this review figures from their work.
\end{acknowledgement}

\bibliographystyle{baaa}
\small
\bibliography{gcarraro}
 
\end{document}